\documentclass[]{aa}
\usepackage{graphicx,psfrag,amsmath}
\usepackage{natbib}
\bibpunct{(}{)}{;}{a}{}{,}
\begin{document}

\title{Vertical distribution of Galactic disk stars
\thanks{Based on observations  made at the Observatoire
        de  Haute  Provence (France). Data only available in electronic 
	form at the CDS (Strasbourg, France}} 
\subtitle{I - Kinematics and metallicity} 

\titlerunning{Kinematics and metallicity}
\author{Soubiran C.\inst{1}, Bienaym\'e O.\inst{2}, Siebert A.\inst{2}}
\offprints{Caroline.Soubiran@observ.u-bordeaux.fr}
%
\institute{Observatoire Aquitain des Sciences de l'Univers,  UMR 5804, 2  
rue de  l'Observatoire,  33270
              Floirac, France  \and Observatoire    astronomique   de   
	      Strasbourg,    UMR   7550,
              Universit\'e   Louis    Pasteur,   Strasbourg,   France}
\date{Received \today / Accepted }
\abstract{Nearly 400 Tycho-2 stars have  been observed in a 720 square
degree field in the direction of the North Galactic Pole with the high
resolution   echelle   spectrograph   ELODIE.   Absolute   magnitudes,
effective   temperatures,  gravities   and  metallicities   have  been
estimated, as well as distances and 3D velocities. Most of these stars
are clump  giants and span  typical distances from 200\,pc  to 800\,pc
to the galactic mid-plane. This  new sample, free of any kinematical
and metallicity bias, is used to investigate the vertical distribution
of  disk stars.   The old  thin disk  and thick  disk  populations are
deconvolved from  the velocity-metallicity distribution  of the sample
and their parameters are determined. The thick disk is found to have a
moderate  rotational lag  of $-51  \pm  5\,\mathrm{km\,s}^{-1}$ with
respect  to  the Sun  with  velocity  ellipsoid $(\sigma_U,  \sigma_V,
\sigma_W)=(63\pm  6, 39\pm  4, 39\pm  4)  \,\mathrm{km\,s}^{-1}$, mean
metallicity   of   [Fe/H]  =$-0.48\,\pm$\,0.05   and   a  high   local
normalization  of 15\,$\pm$\,7\%.   Combining this  NGP sample  with a
local  sample  of  giant  stars  from  the  Hipparcos  catalogue,  the
orientation  of  the  velocity  ellipsoid is
investigated as a  function of distance to the  plane and metallicity.
We  find no  vertex  deviation  for old stars, consistent with an axisymmetric
Galaxy. Paper II  is devoted to the dynamical  analysis of the sample,
puting  new constraints  on the  vertical force  perpendicular  to the
galactic plane and on the total mass density in the galactic plane.

}
\maketitle
%
%
\section{Introduction}
The primary purpose of the observations described in this paper is to
determine the potential  from 300 pc to 1 kpc  from the galactic plane
using an  homogeneous and  complete sample of  red clump giants.  The
measure of the potential close to the galactic plane has been recently
improved thanks  to the Hipparcos observations leading  to an unbiased
estimate        of         the        local        mass        density
$0.076\pm0.015\,$M$_\odot$\,pc$^{-3}$  (based   on  A  and   F  stars,
\citet{CC98a, CC98b}   ,   see   also   \citet{Ph98}   and   \citet{HF00})
constraining the potential  shape up to 125\,pc  from the galactic
plane.   This  result  is  consistent with  previous  and  independent
determinations based on  distant F dwarfs and K  giants at 0.5--1\,kpc
\citep{KG89}.   A direct  implication of these  results is  that our
galactic dark matter component cannot  be strongly flattened and has a
more  or less  spheroidal shape.  In order  to draw  with  an improved
accuracy the potential at various  heights from the galactic plane and
confirm  or not  this property,  new kinematical  data appeared  to be
necessary.\par

For this  task we have  chosen to observe  red clump giants  which are
well suited for galactic studies. The first
and evident interest in studying giants is their intrinsic brightness,
(m$_{\rm   V}\simeq$ 9.3\  at   500\,  pc)   that  allows   to  obtain
high-resolution spectroscopy on  distant stars.  Moreover clump giants
have a  high density in the  solar neighbourhood as  compared to other
intrinsically bright stars, they dominate in the magnitude interval $9
\leq m_{\rm v} \leq 10.5$ and colour interval $0.9 \leq B-V \leq 1.1$.
Thanks to their strong-lined spectra,  K stars are well adapted to the
measure  of  accurate radial  velocities and  atmospheric  parameters:
effective temperature  $T_{\rm eff}$, logarithm of  gravity $\log\, g$
and  metallicity  measured  by  the  iron  abundance [Fe/H].   In
practice,  atmospheric  parameters were deduced from low \mbox{S/N ($\sim20$)}
ELODIE spectra  using the  automated method of  parametrisation TGMET,
based on  a criterion  of minimum distance  to an empirical library  of reference
spectra  \citep{KS98,SK98}.   The same  principle  is used  to
estimate  the  absolute magnitude  of  the  target  stars, leading  to
distances  with an  accuracy  better than  20\%,  unprecedented for  K
giants out of the 250\,pc Hipparcos sphere. \par

This  new  sample  has  been used to determine the vertical force $K_z$ 
and the disk surface mass density as described in Paper II. These data have 
also  provided an  opportunity  to  probe  the
transition  between  the  thin  and  thick disks  and  to  investigate
their properties.  If  the existence of a  thick disk in our  Galaxy is well
established, its  formation and connexion  to the thin disk  are still
matter  of  debate.   Several  models  of  formation,  "top  down"  or
"bottom-up"  are  proposed which  predict  peculiar  features for  the
spatial,  chemical, kinematical  and  age distributions  of the  thick
disk,   including   mean   behaviour   and   dispersions,   gradients,
continuity-discontinuity with other populations. Such predictions are described for
instance in \cite{Ma93}.  Therefore, a detailed knowledge of properties
of the thick disk is necessary to favour one of the proposed models of
formation.  Previous studies of the thick disk properties are numerous
but most  of them  suffer from serious  limitations: local  samples of
selected stars  are biased  in favour of  metal-poor or  high velocity
stars, samples of  tracers (clusters, RR Lyrae...)  are  small and not
necessary representative of the  whole thick disk population, while in
situ  surveys  suffer  from  lower   precision  due  to  the  lack  of
astrometric  and spectroscopic  observations  for faint  stars.  As  a
consequence,  the  parameters of  the  thick  disk  are not  precisely
established.   Its current status  is given  in \cite{N99}.   There is
still a  controversy between the partisans  of a flat  and dense thick
disk, with typical scale height  of 800\,pc and local relative density
of 6-7\% \citep{RR01}, and the partisans of a thick disk with a higher
scale height,  typically 1300\,pc and a lower  local relative density,
of  the order of  2\% \citep{RM93,Ch97}.   Velocity ellipsoid,
asymmetric drift  and mean metallicity are the  other parameters which
characterize the thick disk.  Velocity dispersions are generally found
to  span  typical  values  between 30  and  50  $\mathrm{km\,s}^{-1}$,
sometimes  up  to 80  $\mathrm{km\,s}^{-1}$  in  the radial  direction
\citep{RF89}.   The  asymmetric  drift  ranges  between  -20  and  -80
$\mathrm{km\,s}^{-1}$  and   the  mean  metallicity   from  --0.50  to
--0.80 dex. The view of  the thick disk has been  complicated by the study
of  \cite{MFF90}  who  brought   to  the  fore  low-metallicity  stars
$(-1.6<$[Fe/H]$<-1.0)$ with disk-like kinematics. \cite{CB00} estimate
that 30\% of  the stars with $-1.6<$[Fe/H]$<-1.0$ belong  to the thick
disk  population.   It  remains  unclear whether  this  population  is
separate  from  the thick  disk  or  its  metal-weak tail. Very recently, 
\cite{GWN02} claimed to have observed the relics of a shredded satellite 
forming a thick disk rotating at  100  $\mathrm{km\,s}^{-1}$ at several
kpc from the plane. A  better
understanding  of the thick  disk population(s)  can be  achieved from
accurate distances, 3D velocities  and metallicities for large samples
of  stars, with  no kinematical  or chemical  selection, far  from the
plane where the relative density of  the thick disk is high. Thanks to
our new complete and large sample of K giants at a mean height of 400\,pc above
the plane, the old thin disk and thick disk populations were
investigated with  an unprecedented accuracy  on distances, velocities
and  metallicities.  Combining  metallicities  and  velocities, we
were able  to deconvolve  the thin disk  and thick  disk distributions
providing new estimates of their parameters.\par

The change of the velocity  ellipsoid with respect to height above the
plane and metallicity has also been investigated.  If the stationarity
and  axisymmetry of  the  Galaxy  is generally  considered  as a  good
approximation,  large scale  deviations  produced by  warps, bars,  or
triaxiality of the halo, are also often mentioned.  The orientation of
the velocity ellipsoid is an indicator of such deviations. In the case
of axisymmetry  and stationarity, the velocity ellipsoid  of the oldest stars
is supposed to point towards the  Galactic Center. This is the case in
our  NGP  sample.\par

The  analysis  of the  sample  has  been  carried out  with  different
tools. For the validation of our
methods and for the interpretation of the stellar  content of  our dataset,
we  have used simulations of  the Besan\c con  model of stellar
population  synthesis  of  the  Galaxy \citep{RC86, RRDP02}.
   The   deconvolution  of  the  thin  disk   and  thick  disk
distributions has been performed  in the 4D velocity-metallicity space
using a maximum likelihood algorithm assuming gaussian components.\par

In Sect.\,2  we describe the selection  of the sample  to optimize the
number of  clump giants. Sect.\,3  is devoted to the  determination of
stellar parameters: radial velocities, atmospheric parameters ($T_{\rm
eff}$, $\log\,g$,  $\mathrm{[Fe/H]}$), absolute magnitudes $M_{\textrm
v}$, distances, 3D  velocities with respect to the  Sun.  In Sect.\,4,
observed vertical gradients are compared to simulations of
the  Besan\c con  model, and  the deconvolution  of two  components is
described.  Sect.\,5 deals  with  the vertex deviation. Our conclusions 
are summarized in Sect.\,6.
%
\section{Selection of the sample}
\label{s:selection_sample}

The bulk of  our sample consists of stars selected  out of the Tycho-2
catalogue  \citep{TYC2}.  This  catalogue  provides $V_T$  magnitudes,
$B_T-V_T$  colours,  accurate   proper  motions  and  related  errors.
Tycho-2 photometry can be transformed into Johnson photometry applying
the transformation given in \citet[~chap.~1]{HIP}
\begin{eqnarray}
V=  V_T  -0.090 \, (B_T-V_T) \nonumber \\ 
B-V  = 0.850 \,(B_T-V_T) \,\,\, \mbox{,} 
\end{eqnarray}  

A selection in $B-V$ colour has  been applied to increase the number of
red  clump  stars with  respect  to  dwarfs,  subgiants, RGB  and  AGB
stars. We  have looked  at several studies,  based on  Hipparcos data,
which use and discuss the properties  of the position of the red clump
in the  HR diagram,  used for instance  as distance indicator  for the
galactic      center     \citep{Pa98}      or      nearby     galaxies
\citep{GS01}.  According to these  studies, we  have chosen the colour
interval 0.9\,$<B-V<$\,1.1  that optimizes the detection  of red clump
stars.  The  lower $B-V$  limit  at  0.9  rejects the  brightest  main
sequence  stars  ($M_{\textrm v}<5$),  some  subgiants  and also  most
giants  of the  blue horizontal  branch  (the "clump  stars" with  the
lowest metallicities).  The  upper $B-V$ limit at 1.1  rejects most of
RGB and  AGB stars. We have set  the limiting magnitude at  V=10.5 to limit
the exposure time of the spectroscopic observations, also because dwarfs
begin to  be dominant at fainter magnitude.  Observing conditions have
finally lowered the limiting magnitude of the sample.\par

In order  to have a regular  distribution of stars  versus distance to
the galactic  plane, the survey has  been carried out  on two circular
fields towards the  NGP.  The two observed fields  have different size
and  limiting apparent magnitude (Table~\ref{t:field}).
The  first one is  shifted by  10 degrees  from the  NGP to  avoid the
region  of  the  Coma--Berenices open  cluster  ($\ell\,=\,221\degr,\,
b\,=\,+84\degr$). A  circular region  of $4.5\degr$ radius  around the
center of the  cluster has been removed from  the second field centred
on the NGP.   As the two fields overlap, the  common region is defined
as belonging  to the deepest field.   Some stars were  excluded {\it a
priori} from the observations: i) stars identified as dwarfs thanks to
their accurate  Hipparcos parallaxe, ii) stars  identified as variable
or multiple system.\par

A  total of  387  stars have  been  observed with  the Elodie  echelle
spectrograph at OHP, 8 of  them have been observed twice. According to
individual errors  quoted in the  Tycho-2 catalogue, the sample  has a
mean error  of $1.3\,  \mathrm{mas\,yr}^{-1}$ on proper  motions, 
corresponding to an error  of $3.25\, \mathrm{km\,s}^{-1}$ on tangential
velocities at 500\,pc.
The characteristics  of   the  two  fields  are  summarized  in
Table~\ref{t:field}.
\begin{table*}[hbtp]
\caption{
\label{t:field}
Main characteritics of the two  observed fields, ${\it l}_c$ and ${\it
b}_c$ are the galactic coordinate of the center of the fields}
\begin{center}
\begin{tabular}{l c c c c c c}
\hline \hline  & ${\it  l}_c$ & ${\it  b}_c$ & limiting  magnitude &
interval of completeness &  number of & Surface\\ & (deg) & (deg) &in
V$_J$ mag  & in V$_J$  mag & stars  observed & square  deg\\ 
\hline 
Field  1 & 35.46 & 80. & 10.48  &  7.2 - 10.1 & 246  & 309.4\\ 
Field 2  & 0.    & 90. &  9.57  &  7.2 - 9.2  & 141  & 410.1$^*$\\ 
\hline 
\multicolumn{7}{l}{$^*$
this surface is  corrected for the overlap region and for  the presence of the
Coma Berenices open cluster}
\end{tabular}
\end{center}
\end{table*}
%
\section{Stellar parameters}
\subsection{Spectroscopic observations, radial velocities}
The observations were carried out with the echelle spectrograph Elodie
on the  1.93\,m-telescope at the  Observatoire de Haute  Provence. The
performances  of this  instrument  are described  in \citet{BQ96}.   A
total of 22 nights was  attributed to this program in March-April 2000
and  April-May 2001.   The  exposure time  on  individual targets 
ranges from 90 seconds to 35  minutes, with a mean value of 9 minutes
and a total  of 61 hours.  The resulting spectra  cover the full range
390 -- 680 nm at a resolving power  of 42\,000.  Spectrum extraction,
wavelength calibration and measurement of radial velocities
have been performed at the  telescope with the on-line  data reduction
software.       The radial velocity accuracy is better than
\mbox{$1\,\mathrm{km\,s}^{-1}$}  for the  considered stars  (K stars).
Our    sample   spans    radial   velocities    from   --92    to   85
$\mathrm{km\,s}^{-1}$    with     a    mean    value     of    $-13.5$
$\mathrm{km\,s}^{-1}$. The mean S/N of the spectra at 550 nm is 23. 
\subsection{Atmospheric parameters ($T_{\rm eff}$, $\log\,g$, $\mathrm{[Fe/H]}$) 
and absolute magnitudes $M_{\rm v}$}
\subsubsection{TGMET and its library of reference spectra}

 Atmospheric parameters and absolute magnitudes were obtained with the
TGMET  software  \citep{KS98}.  TGMET  is  a  minimum distance  method
(reduced  $\chi^2$ minimisation)  which measures  similarities between
spectra in  a quantitative way and  finds for a  given target spectrum
the most  closely matching template  spectra in a library.   The TGMET
library was built from  the ELODIE database described in \citet{PS01},
selecting  high  S/N spectra  of  reference  stars  having well  known
atmospheric parameters from published  detailed analyses listed in the
Catalogue  of [Fe/H]  determinations \citep{CS01}.   The  selection of
reference  stars was extended  to stars  with published  $T_{\rm eff}$
from  the lists  of \citet{BLG98,dB98,AA96a,AA99a}  and to  stars with
available  $V-K$  colour  index  in  the  catalogue  of  \citet{MM78},
calibrated    into    $T_{\rm   eff}$    using    the   formulae    of
\citet{AA96b,AA99b}.     Multiple   determinations    of   atmospheric
parameters  for the  same reference  star were  averaged,  giving more
weight to the most recent analyses.\par

Thanks to  TGMET, absolute  magnitudes $M_{\textrm v}$  were estimated
simultaneously with the atmospheric parameters, based on the fact that
stars    having     similar    spectra    have     similar    absolute
magnitudes. Therefore, the TGMET  reference library was completed with
Hipparcos  stars having parallaxes  with a  relative error  lower than
30\%.  Their absolute magnitudes $M_{\textrm v}$ were derived from the
Tycho-2 $V_T$ apparent magnitude, transformed into $V$ Johnson band, and
these  stars  were  used  as  reference stars  for  $M_{\textrm v}$.  
Finally  the TGMET  library which was  used in  this study
included 1112 spectra  of FGK stars, 598 with  available $T_{\rm eff}$
from the literature, 558 with available $\log\,g$, 577 with available
[Fe/H], 1079 with available  $M_{\textrm v}$ determined from Hipparcos
(884  from a parallax  with a  relative error  lower than  10\%).  The
parameters ($T_{\rm  eff}$, $\log\,g$, [Fe/H], $M_{\textrm  v}$) of a
target star processed  by TGMET are given by the  weighted mean of the
parameters  of  the  best  matching reference  spectra. 

\subsubsection{TGMET external accuracy} 

The external accuracy of TGMET depends on the quality of the
reference system defined by the library of spectra. Ideally the reference library
should be a perfect grid, paving regularly the parameter space. In practice, an
observed library does not describe all parts of the parameter space with the same 
sampling. For instance, very metal-poor stars  or cool dwarfs  are less represented  
in the library and results of lower  quality are expected  for this class  of stars.
Moreover, the reference spectra being parametrized from the literature (and Hipparcos), some
errors affect their parameters (typical errors  quoted in  detailed spectral  
analyses of the literature range from 
50 K to 150  K for $T_{\rm eff}$, 0.1 to  0.3 for $\log g$, and 0.05 to  0.1 for [Fe/H]).
\par

\begin{figure}[hbtp]
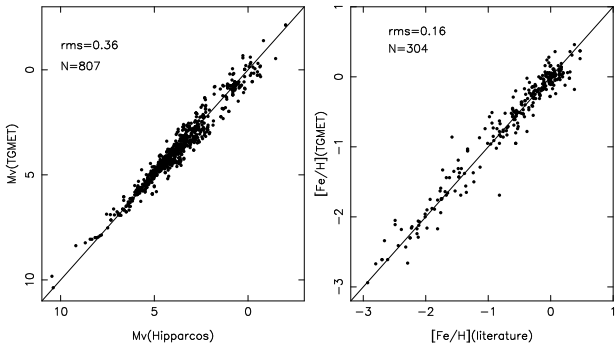
                                 
    \begin{center}
    \includegraphics[width=4cm]{ps/bib_Mv.ps}     
    \includegraphics[width=4cm]{ps/bib_FeH.ps}     
    \caption{Comparison of literature metallicities and Hipparcos  
    absolute magnitudes to their TGMET counterparts for a subset of high
    quality reference stars. }  
    \label{f:bib_FehMv} 
    \end{center} 
    \end{figure}

The quality of the library was assessed, and simultaneously the global accuracy of  
the TGMET atmospheric  parameters and absolute  magnitudes was
estimated. We have tested the ability of TGMET to recover the parameters of stars considered 
as highly reliable reference stars. For this task we have selected in the library 
only stars with  the smallest error bars on their external parameters, having 
several determinations of atmospheric  
parameters in agreement in the literature and a relative parallax error lower than 10\% in Hipparcos.
For each star of this subset, the corresponding spectrum was removed from the library, its
TGMET parameters were estimated with  respect to  the rest  of the library,  and compared 
to the literature and Hipparcos ones, as shown in 
Fig.~\ref{f:bib_FehMv} for [Fe/H] and Mv. Differences  between  TGMET  and   literature  
parameters  have a standard deviation of  0.16  in  [Fe/H],  
and  0.36  in $M_{\textrm v}$. The  standard deviation in $M_{\textrm v}$ 
is lower for
dwarfs than for giants : 0.25 for dwarfs ($M_{\textrm v} > 4$),
0.37 for subgiants ($3 < M_{\textrm v} \le 4$), 0.32 for clump giants 
($0.5 < M_{\textrm v} < 1$), 0.47 for the other giants. Good results  for
clump giants are explained by the number of
reference spectra in the library for this type of stars, with accurate Hipparcos parallaxes.\par

\begin{figure}[hbtp]                                 
    \begin{center}
    \includegraphics[width=4cm]{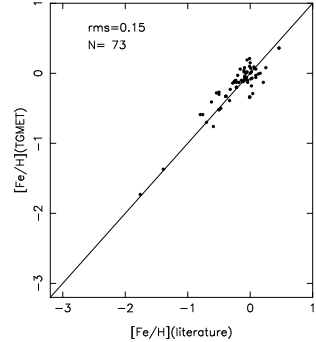}     
    \caption{Comparison of literature and TGMET metallicities for the 73 reference 
    stars which were effectively used to determine the metallicity of the target stars.}  
    \label{f:bib_Feh_util} 
    \end{center} 
    \end{figure}

In
practice only a fraction of the library was useful for TGMET to determine 
the parameters of the NGP
targets : a subset of 117 reference spectra was selected on the basis of their
similarity with the targets. 
These stars have parallax errors  ranging from 0.45 to 1.7 mas, with a median of 0.8 mas.
The median relative error on parallax of 3.45\% for this subset ensures the high quality
of our reference system for $M_{\textrm v}$. Several (16) stars with
a relative error greater than 10\% correspond to metal poor
giants which have to be kept in the library to represent the low metallicity population. The
iron abundance of the target stars was determined in practice from 73 reference stars. 
Their literature and TGMET metallicities are compared in Fig.~\ref{f:bib_Feh_util}. On average 
the TGMET metallicities are overestimated by 0.06 dex with an rms of 0.15.   \par


Unfortunatly the absolute magnitudes from Hipparcos
parallaxes are affected by the Lutz-Kelker bias, especially among giants. This causes
an additional external error which must be taken into account. In paper II, 
the Lutz-Kelker bias is estimated for a local sample of clump giants. It is shown
that on average their absolute magnitudes  have to be corrected by -0.09 mag
with an rms of 0.13 mag. We have not attempted to correct individual absolute magnitudes of the
reference stars because the Lutz-Kelker bias is estimated from the luminosity function
of the parent population which is unknown for most giants of the library, except for the clump ones.
Combining the different sources of external errors,  we estimate the TGMET absolute magnitudes 
of clump giants to be affected by a mean error of 0.35
mag, corresponding to an error in distance of 17\%, with a systematic overestimation of distances 
by 4\%. 

\subsubsection{TGMET internal precision} 
We have investigated the  internal precision of TGMET  
by comparing the parameters obtained on target stars in the case of double observations. 
 The TGMET parameters of the 8 target stars 
observed twice are given in Table~\ref{t:tgmet_bis}. 
Both solutions show a
very  good  agreement. The results are however of lower quality for  T2012-00935,  
because  one of  the exposures has the lowest S/N of our observations (S/N=5), at the limit of 
the possibilities of TGMET, and
for T2010-01190 because its spectrum shows an enlarged profile (FWHM =
$15.6\, {\rm km\,s}^{-1} $) probably indicating a spectroscopic binary (enlarged profiles 
 with FWHM $> 14\, {\rm km\,s}^{-1} $ were only observed for 6 target stars, 2 double lined spectroscopic
 binaries were detected).
Excluding these 2 non-representative cases, we compute rms differences of 0.02 dex in metallicity 
and 0.075 in absolute magnitude, giving an estimate of the precision of
the TGMET method.\par
\begin{table*}[hbtp]
\caption{
\label{t:tgmet_bis}
Comparison  of  the  parameters   obtained  in  the  case  of  double
observations.  The  heliocentric radial velocity  $V_r$ was obtained  with the
on-line cross-correlation software, together  with the FWHM which measures the
true resolution of the spectrum. The atmospheric parameters 
($T_{\rm eff}$, $\log g$, $\mathrm{[Fe/H]}$) and absolute magnitudes $M_{\textrm v}$
have been obtained with the TGMET software.}
\begin{center}
\begin{tabular}{c c c c c c c c}
\hline \hline  Tycho Id &  S/N & $Vr  $ & FWHM  & $T_{\rm eff}$  & $\log g$ &  [Fe/H] &
$M_{\textrm v}$ \\ & at 550 nm & $\mathrm{km\,s}^{-1}$ & $\mathrm{km\,s}^{-1}$
& K & & & \\ \hline T1461-00774 & 15.8  & 16.55 & 11.2 & 4924 & 2.74 & -0.23 &
0.756 \\ T1461-00774  & 15.8 & 16.50 & 11.2  & 4929 & 2.75 &  -0.20 & 0.598 \\
\hline T1462-00831  & 11.9  & 1.26  & 10.7  & 4807 &  3.26 &  0.05 &  2.077 \\
T1462-00831 &  21.8 &  0.89 &  11.0 &  4807 & 3.26  & 0.05  & 2.078  \\ \hline
T1987-00205  &  17.5 &  -27.53  & 11.3  &  4776  & 2.59  &  -0.20  & 0.805  \\
T1987-00205 &  20.4 & -27.56 &  11.2 & 4765 &  2.50 & -0.19 &  0.781 \\ \hline
T2007-01215 & 10.5 & 0.24 & 12.6 &  4974 & 2.82 & -0.43 & 0.647 \\ T2007-01215
& 23.3  & 0.09 & 12.5 &  4888 & 2.57 &  -0.48 & 0.689 \\  \hline T2010-01190 &
11.5 &  -0.02 & 15.6  & 4722 & 2.62  & -0.04 &  0.678 \\ T2010-01190 &  38.0 &
-0.08 & 15.6 & 4895 & 3.14 & 0.15 & 1.807 \\ \hline T2012-00935 & 5.0 & 5.45 &
11.4 & 5273 & 3.56 & 0.03 & 3.087 \\ T2012-00935 & 40.3 & 5.22 & 11.1 & 4918 &
3.03 &  -0.07 & 2.136 \\  \hline T2527-01831 & 30.4  & -28.72 & 11.3  & 4801 &
2.52 &  -0.02 & 0.969 \\ T2527-01831  & 17.5 & -28.82  & 11.3 & 4804  & 2.51 &
-0.03 &  0.953 \\ \hline T2532-00687  & 21.3 & -66.09  & 10.9 & 4711  & 2.53 &
-0.53 &  1.329 \\ T2532-00687 & 7.9  & -66.99 & 11.0  & 4709 & 2.46  & -0.54 &
1.202 \\ \hline
\end{tabular}
\end{center}
\end{table*}

\subsubsection{The ([Fe/H],  $M_{\textrm v}$) distribution of the NGP sample}
\begin{figure}[hbtp]                                 
    \begin{center}
    \includegraphics[width=6cm]{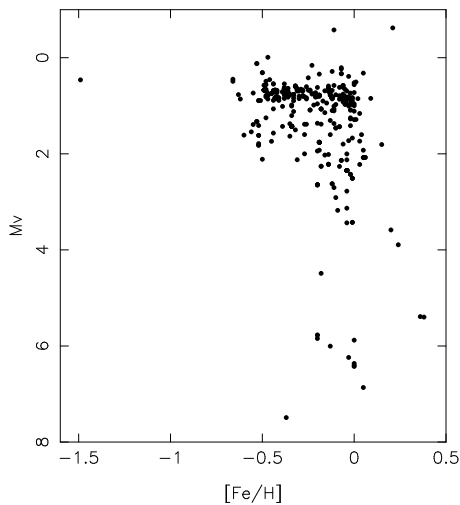}  
    \caption{Metallicity and
    absolute magnitude  distribution of the NGP sample  (387 stars).}  
    \label{f:FehMv} 
    \end{center} 
    \end{figure}

The parameters obtained with TGMET for  the NGP targets range from 4322 K to 5698 K 
in $T_{\rm eff}$, from 1.02
to 4.70 in   $\log\,g$, from -1.49 to +0.38 in [Fe/H], from -0.621 to
7.490 in $M_{\textrm v}$. The most  important parameters for this study 
are the  metallicities and absolute
magnitudes which  lead to individual  distances. Figure~\ref{f:FehMv} represents
 the  NGP sample  in the  plane ([Fe/H],  $M_{\textrm v}$). Some comments have to
  be done about this plot:
\begin{itemize}
\item the sample spans the full range of luminosities, from dwarfs to bright giants
\item the clump is clearly identified as an overdensity  around $M_{\textrm v}$=0.81. 
\item the  selection of  giants  was efficient : dwarfs or subgiants ($M_{\textrm v}\,>\,3$)
represent only 6\% of the sample
\item there  is no  dependency of  the absolute magnitude  of clump  giants on
[Fe/H]
\item there is a cut in the metallicity distribution at [Fe/H]=--0.65 with only
one metal-poor star at [Fe/H]=--1.49.
\end{itemize}

 We have checked if the lack of stars with [Fe/H] $< -0.65$ could be  due to  
the distribution  of  the TGMET
reference stars  in the  parameter space. According to Fig.~\ref{f:bib_Feh_util}, 
several reference stars in the range $-1.00 <$ [Fe/H] $< -0.65$ exhibit similarities
with the targets, but the nearest neighbours are more metal-rich. Several clump giants with
lower metallicity exist in the library (HD166161, HD175305, HD044007 for instance) but they
have not been
selected by TGMET to present similarities with the targets. We conclude that TGMET and its
library cannot explain  by their own the complete 
absence of target stars in this metallicity range, even if an overestimation of [Fe/H] by
0.06 was established (Sect. 3.2.2).\par

Basic data from Tycho2 and TGMET parameters of the 387 NGP targets are given in a table only 
available in electronic form at the CDS (http://cdsweb.u-strasbg.fr/Cats.html).

\subsection{Distances and spatial velocities}
\begin{figure}[hbtp]
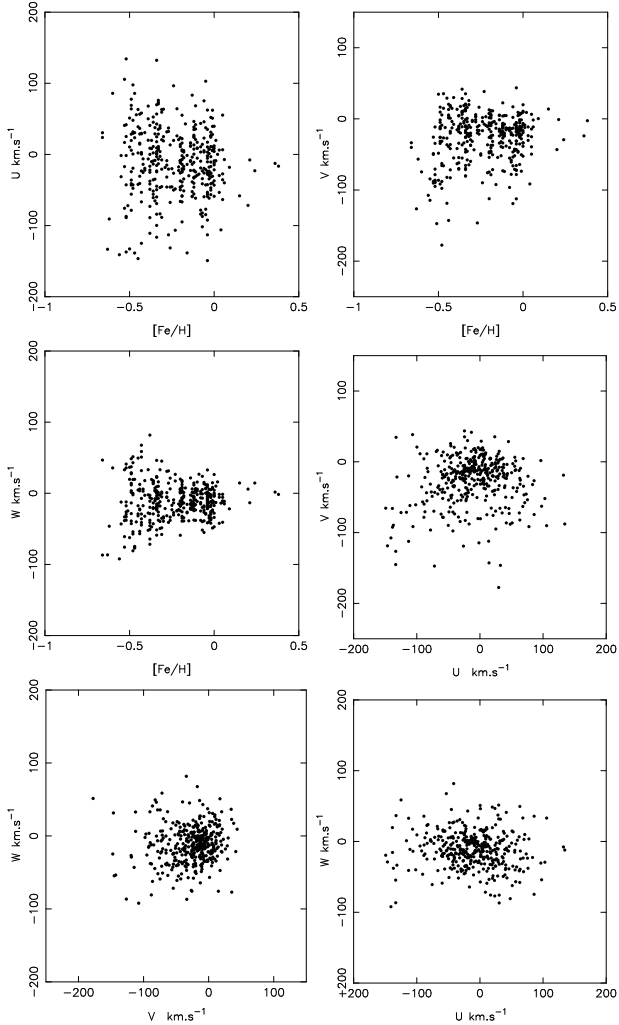
  
    \centering  
    \includegraphics[width=4.cm]{ps/FU.ps}
      \includegraphics[width=4.cm]{ps/FV.ps}
      \includegraphics[width=4.cm]{ps/FW.ps}
      \includegraphics[width=4.cm]{ps/UV.ps}
      \includegraphics[width=4.cm]{ps/VW.ps}
      \includegraphics[width=4.cm]{ps/UW.ps}  
      \caption{Distributions  of  the
       NGP sample  (387 stars)  in  metallicity and  3D velocities  with
      respect to the Sun.} 
      \label{f:fehUVW}
      \end{figure}

Distances  have  been  computed  for  all  the target  stars  from  the  TGMET
$M_{\textrm  v}$ and  Tycho-2  $V_T$  magnitude transformed  into  Johnson $V$ by
relation 1.  No
correction of interstellar  absorption was applied since it  is supposed to be
very  low in  the NGP  direction.
Proper motions,  distances and  radial velocities  have  been combined
through the  equations of  \citet{JS87} to compute  the 3  velocity components
(U,V,W) with  respect to the Sun (the U axis points towards the Galactic Center).  
Figure~\ref{f:fehUVW}  shows the metallicity and  velocity distributions of the whole sample
in  different combinations.   For a  sake of
clarity, we  have excluded  from these plots  2 stars with  extreme parameters
which   may   belong   to    the   halo: T2004-00702 ([Fe/H]\,=\,--1.49, z\,=\,755\,pc,
U\,=\,--133.4\,$\mathrm{km\,s}^{-1}$, V\,=\,--145.1\,$\mathrm{km\,s}^{-1}$, 
W\,=\,--54.6\,$\mathrm{km\,s}^{-1}$)  and T2524-01735 ([Fe/H]\,=\,--0.38, z\,=\,475\,pc,
U\,=\,40.8\,$\mathrm{km\,s}^{-1}$, V\,=\,--280.0\,$\mathrm{km\,s}^{-1}$, 
W\,=\,-37.8\,$\mathrm{km\,s}^{-1}$).
From these plots, it can be seen that stars with high radial motion ($U < -100\,
\mathrm{km\,s}^{-1}$) or  slow rotation ($V <  -80 \,\mathrm{km\,s}^{-1}$) exist
at   all  metallicities.   The   velocity  dispersions   are  higher   at  low
metallicities, especially for $W$. Here  we recall that  our W
velocities are of an unprecedented accuracy for a survey at this mean distance
as the contribution of the ELODIE accurate radial  velocity to W is higher than 94\% in
the considered direction. 

\section{Distances, abundances and kinematics: stellar components}
\subsection{Expected relative densities}

The median distance of the sample  to the galactic plane is 400\,pc. We expect
to find at this distance a  mixture of disk components, mainly the old
thin disk  and the  thick disk.  For instance, adopting exponential  laws to
modelise their  vertical density, with a  scale height of 300\,pc  for the older
thin  disk,  800\,pc  for  the  thick  disk with  a  local  density  of  6.2\%
\citep{RR01}, proportions of 9, 13, 18 and 25\% of thick disk stars are expected
at mean distances of 200, 400, 600 and 800\,pc respectively.  For a scale height
of 1400\,pc and  a local density of 2\% \citep{RM93}  the proportions of thick
disk stars fall to 3, 5, 8 and 13\%. With a local relative density well below 1\%,
the halo population is expected to be marginally represented in the NGP sample. 
The aim  of this section  is to identify disk components in the NGP sample,  
investigate their behaviour up to 800\,pc from the plane and 
determine their mean parameters. 

\subsection{Comparison with the  Besan\c  con  model}
  The NGP sample was compared with several samples of pseudo-stars simulated with
the most recent version of the  Besan\c  con  model of population synthesis \citep{RRDP02}.
 This model is based on
a semi-empirical approach, involving physical constraints and current knowledge of the formation and 
evolution of the Galaxy. For observing  conditions given as inputs
(direction and size  of  the fields, cuts in apparent magnitude and colour, observational errors 
on apparent magnitudes, proper motions and radial velocities), the model
returns a list of pseudo-stars with their fundamental properties (distance, absolute magnitude, age,
3D velocity, metallicity) and the corresponding observables affected by errors. The fundamental 
properties coming from theoretical distributions,  we had
to degrade metallicities and absolute magnitudes with the typical errors of the NGP sample to 
mimic
the real data. Metallicities were degraded with a gaussian error of 0.16 dex
and absolute magnitudes with an error varying from 0.25 for dwarfs to 0.35 for clump giants
and 0.47 for other giants (see Sect. 3.2.1). Then distances and space velocities were 
re-computed from
degraded absolute magnitudes combined to simulated apparent magnitudes, proper motions and 
radial velocities.\par
 
The Besan\c  con  model involves 7 thin disk  components with mean  ages from 0.15  to 10\,Gyr,
a thick disk component, a halo and a bulge. We
have chosen to test our data against 2 different models: one involving a metal-poor, kinematically
hot thick disk (hereafter model 1) and another one less metal deficient and kinematically colder 
(hereafter model 2). The corresponding parameters used as inputs for the thin disk and the 
thick disk in
both models are given in Table~\ref{t:thickdisk}. In both cases, the
scale height of the thick disk is 800\,pc and its relative density in the solar neighbourhood is
6.2\%. The asymetric drift of the different components with respect to the Sun ($V_{\rm lag}$) is 
based on 
$V_{\rm LSR}=220\,\mathrm{km\,s}^{-1}$ and $V_\odot=6\,\mathrm{km\,s}^{-1}$.
 For each model, 3 simulations with our observational conditions have been performed. We have 
 obtained
422, 454, 433 pseudo-stars with model 1 and 395, 434, 422 pseudo-stars with model 2. These 
numbers
have to be compared with the 387 stars of the NGP sample. According to the fact
that the NGP sample is not complete down the limiting magnitude, and that several dwarfs and 
double stars
were eliminated {\it a priori}, the model counts agree very well with the observations.\par
\begin{figure}[hbtp]
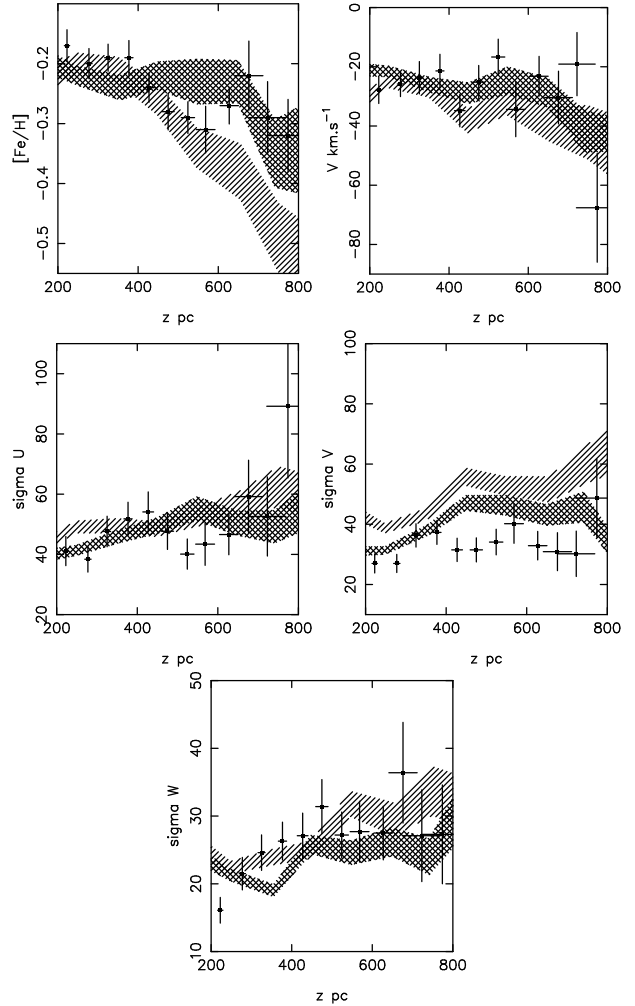
  \centering  
      \includegraphics[width=4.cm]{ps/zf.ps}
      \includegraphics[width=4.cm]{ps/zv.ps}
      \includegraphics[width=4.cm]{ps/zsu.ps}
      \includegraphics[width=4.cm]{ps/zsv.ps}
      \includegraphics[width=4.cm]{ps/zsw.ps}  
      \caption{Vertical gradients of
      [Fe/H],$  V,  \sigma_U,  \sigma_V,  \sigma_W$ with  standard error
      bars.  Dots:  the NGP  sample, hatched area: model 1, double hatched: model 2.} 
      \label{f:zgrad} 
      \end{figure}

\begin{table*}[hbtp]
\caption{
\label{t:thickdisk}
Inputs for the old thin disk and thick disk in the Besan\c con  model.}
\begin{center}
\begin{tabular}{  l  c  c  c  c  c  c  c  c }
\hline 
\hline 
    & $\mathrm{[Fe/H]}$ & $  \sigma_{\rm [Fe/H]} $& $V_{\rm lag}  $&$\sigma_U $ &$\sigma_V $ 
    &$\sigma_W $ \\
    &  & & $\mathrm{km\,s}^{-1}$&$\mathrm{km\,s}^{-1}$&$\mathrm{km\,s}^{-1}$&$\mathrm{km\,s}^{-1}$&\\
\hline
thin disk 5 Gyr    & -0.07 & 0.18 & -17 & 37 & 24 & 15 \\
thin disk 7 Gyr    & -0.14 & 0.17 & -21 & 43 & 28 & 18 \\
thin disk 10 Gyr   & -0.37 & 0.20 & -21 & 43 & 28 & 18 \\
\hline 
thick disk model 1 & -0.70 & 0.30 & -86 & 80 & 60 & 55 \\
thick disk model 2 & -0.48 & 0.30 & -56 & 60 & 45 & 40 \\
\hline 
\end{tabular}
\end{center}
\end{table*}    
We compare  in Fig.~\ref{f:zgrad} the vertical gradients  of [Fe/H], $V,
\sigma_U, \sigma_V, \sigma_W$ which have  been obtained by a simple statistic
in several bins of distances in the NGP sample, and in the 
simulated ones which were merged to improve the statistics. In general, the data 
show a better agreement with model 2 than
with model 1. But the error bars on the observed and simulated data are such that
it is not possible to favour one of the 2 models from only these vertical gardients.
In particular, the V dispersion in  the NGP sample is much lower than in both models.\par

\subsection{Deconvolution into gaussian components}
\subsubsection{Stochastic Estimation Maximisation}

In an attempt to identify more precisely the contribution of the thin disk and
thick disk  in the NGP sample, we have  performed the deconvolution of these 2  components.
We have used the SEM algorithm \citep{CD86}, a noninformative method
which  solves  iteratively  the  maximum  likelihood  equations,  with  a
stochastic   step,  in   the  case   of  multivariate   mixture   of  gaussian
distributions.  This  method has been previously used  to deconvolve the
thin and thick disk populations from velocity distributions \citep{So93,OB96}.
The algorithm is started  with an
initial  position  thrown  at  random,  assuming that the sample is a mixture of 2 or 3 discrete 
 components. 
Each deconvolution is repeated 150 times. Most of the time the algorithm converges the 
150 times to the same solution. In case of multiple solutions, the most frequent one is adopted. 
The  combinations of variables  ([Fe/H],U,V,W) and  (U,V,W) were used.
Two bins  of  distance were considered  (200 -- 400\,pc  and  400 -- 800\,pc) to 
take into account the change of relative densities of the components at various distances 
to the plane.\par

\subsubsection{Testing SEM on simulated data}
Provided that the simulated data represents a mixture of identified components, we have tested 
the ability of SEM to recover their undelying metallicity and velocity distributions.
The SEM outputs are the relative density, mean and standard deviation of 
(U, V, W) and eventually [Fe/H] of each component to be compared to the model inputs. It is also possible 
to examine the ratios and orientation 
of principal axis of the velocity ellipsoid of each component but large error bars
were obtained on these parameters which are not considered in this section.  We have obtained 
48 solutions with the simulated data (2 bins of distance, 2 models, 3 simulations per model, 3D and
4D deconvolution, 2 or 3 components assumed as initial condition). The deconvolution into 3 components was
successful in  very few cases,  with a solution not physically realistic,
one  component having a  very low standard deviation. Therefore we have only considered the 
deconvolutions into 2 components, all successful and stable. The corresponding
averaged  parameters are presented in  Table~\ref{t:semMod}, with their standard deviation.
As expected, kinematical parameters and mean metallicities obtained for the thin disk in the 2 bins of 
distance are in the range of the Besan\c con model inputs (Table~\ref{t:thickdisk}). 
However the relative 
density of
the thin disk is found to be lower than expected in the first bin of distance. According to the 
density laws of the thin and thick disks adopted as inputs for the model, 87\% of pseudostars 
should belong to the thin disk at 400\,pc from the plane.
On the other hand, we obtain a large error bar on this parameter, and the value of 78$\pm$10\% agrees
within $1\sigma$ with the input of the model.\par

The error bars obtained for the thick disk are quite large and illustrate the difficulty to
deconvolve overlapping populations. The parameters of the deconvolved thick disk generally agree
within $1\sigma$ with the model parameters in Table~\ref{t:thickdisk}. For several parameters the 
agreement is within $2\sigma$.
We conclude that
SEM is able to recover the mean parameters of the thin disk and the thick disk, even if they 
strongly overlap as in 
model 2, using 3D or 4D small samples ($N< 200$) affected by observational errors.\par
\begin{table*}[hbtp]
\caption{
\label{t:semMod}
SEM deconvolution on the simulated data. The mean parameters are given
with their standard deviation.}
\begin{center}
\begin{tabular}{  l  c c  c  c  c  c  c  c  c }
\hline 
\hline 
   & \% & $\mathrm{[Fe/H]}$ & $  \sigma_{\rm [Fe/H]} $& $V_{\rm lag}  $&$\sigma_U $ &$\sigma_V $ 
    &$\sigma_W $ \\
    & &  & & $\mathrm{km\,s}^{-1}$&$\mathrm{km\,s}^{-1}$&$\mathrm{km\,s}^{-1}$&$\mathrm{km\,s}^{-1}$&\\
\hline
thin disk 200-400 pc&78$\pm$10 &-0.16$\pm$0.08&0.23$\pm$0.03& -16$\pm$2&37$\pm$4&24$\pm$3&17$\pm$4 \\
thin disk 400-800 pc&75$\pm$6 &-0.17$\pm$0.05&0.29$\pm$0.02&-15$\pm$6&40$\pm$4&26$\pm$3&16$\pm$2 \\
\hline 
thick disk model 1 200-400 pc&21$\pm$8&-0.58$\pm$0.02&0.27$\pm$0.02&-70$\pm$10&78$\pm$14&59$\pm$5&41$\pm$7 \\
thick disk model 1 400-800 pc&24$\pm$5&-0.71$\pm$0.03&0.32$\pm$0.02&-97$\pm$3 &73$\pm$6&73$\pm$11&50$\pm$4 \\
thick disk model 2 200-400 pc&19$\pm$7 &-0.50$\pm$0.10&0.30$\pm$0.05&-67$\pm$9&56$\pm$11&44$\pm$7&29$\pm$3 \\
thick disk model 2 400-800 pc&31$\pm$3&-0.42$\pm$0.07&0.33$\pm$0.01&-65$\pm$11&70$\pm$7&43$\pm$7&35$\pm$7 \\
\hline 
\end{tabular}
\end{center}
\end{table*}    

\subsubsection{The thin and thick disks in the NGP sample}
After  these tests  on  simulated data,  SEM  was applied  to the  NGP
sample. Each bin of distance includes 166 stars. We had to remove from
the second bin, 2 outlier stars previously mentionned in Sect. 3.3. 
Results are presented in Table~\ref{t:semNGP} in the 4D and 3D
cases.\par 
\begin{table*}[hbtp]
\caption{
\label{t:semNGP}
4D and 3D SEM  deconvolution into 2  gaussian components for the  observed
NGP sample, in 2 bins of distance. The parameters are given with their standard errors.}
\begin{center}
\begin{tabular}{  l  c c  c  c  c  c  c  c  c }
\hline 
\hline 
   & \% & $\mathrm{[Fe/H]}$ & $  \sigma_{\rm [Fe/H]} $& $V_{\rm lag}  $&$\sigma_U $ &$\sigma_V $ 
    &$\sigma_W $ \\
    &  &  & & $\mathrm{km\,s}^{-1}$&$\mathrm{km\,s}^{-1}$&$\mathrm{km\,s}^{-1}$&$\mathrm{km\,s}^{-1}$&\\
\hline
thin disk 200-400 pc 3D&77$\pm$7 &       --     &     --      & -11$\pm$2&37$\pm$2&17$\pm$1&20$\pm$1 \\
thin disk 200-400 pc 4D&75$\pm$7 &-0.16$\pm$0.01&0.23$\pm$0.03& -11$\pm$2&36$\pm$2&17$\pm$1&19$\pm$1 \\
thin disk 400-800 pc 3D&60$\pm$6 &       --     &     --      &-8$\pm$3&39$\pm$3&19$\pm$1&23$\pm$2 \\
thin disk 400-800 pc 4D&65$\pm$6 &-0.19$\pm$0.01&0.29$\pm$0.02&-20$\pm$3&44$\pm$3&28$\pm$2&19$\pm$1 \\
\hline 
thick disk 200-400 pc 3D&23$\pm$4&       --     &     --      &-69$\pm$5&68$\pm$8&33$\pm$4&28$\pm$3 \\
thick disk 200-400 pc 4D&25$\pm$4&-0.27$\pm$0.03&0.27$\pm$0.02&-65$\pm$5&67$\pm$7&34$\pm$4&28$\pm$3 \\
thick disk 400-800 pc 3D&41$\pm$5&       --     &     --      &-59$\pm$4&64$\pm$6&34$\pm$3&36$\pm$3 \\
thick disk 400-800 pc 4D&35$\pm$5&-0.42$\pm$0.01&0.32$\pm$0.02&-44$\pm$6&62$\pm$6&44$\pm$4&42$\pm$4 \\
\hline 
\end{tabular}
\end{center}
\end{table*}    

The success of the  deconvolution of the sample
into 2  discrete components supports the  notion of a thick  disk as a
separate population, by  opposition to being the tail  of a continuous
disk.  Together with  the existence of vertical gradients,  this is of
importance  to constrain  models.   \par

In the nearest bin of distance, all the  kinematical
parameters  obtained with  the 3D  and 4D  deconvolution  agree within
$1\sigma$,  showing that [Fe/H] is less discriminant than
velocities to  separate the disk populations. In the 400 -- 800 pc  
bin, a significant difference is observed in V when [Fe/H] is used
in the deconvolution. However, the agreement being still within
$2\sigma$, we adopt as the final parameters for the thin disk the average
of the 4 solutions (Table~\ref{t:defparam}). The velocity ellipsoid
($\sigma_U, \sigma_V, \sigma_W)= (39\pm2, 20\pm2, 20\pm1)\,\mathrm{km\,s}^{-1}$ is
typical of the old thin disk population.   In their study of local
kinematics  from   Hipparcos  data,  Dehnen \& Binney  (1998) and Bianaym\'e (1999) have
determined dispersions  for the oldest thin disk stars in the range
34 -- 38$\,\mathrm{km\,s}^{-1}$ in U, 23 -- 25\,$\mathrm{km\,s}^{-1}$ in V,
17 -- 21$\,\mathrm{km\,s}^{-1}$ in W. From stars at larger distances, we obtain
similar values, showing that no significant vertical gradient is present in
velocity dispersions of the thin disk.  \par

\begin{table*}[hbtp]
\caption{
\label{t:defparam}
Adopted parameters for the thin and thick disks from the deconvolution of the NGP sample. The 
relative densities are given for z=0. The error bars were estimated from the dispersion
of the SEM solutions on real and simulated data.}
\begin{center}
\begin{tabular}{l c c c c c c c c c }
\hline 
\hline 
   & \% & $\mathrm{[Fe/H]}$ & $  \sigma_{\rm [Fe/H]} $& $V_{\rm lag}  $&$\sigma_U $ &$\sigma_V $ 
    &$\sigma_W $ \\
    &  &  & & $\mathrm{km\,s}^{-1}$&$\mathrm{km\,s}^{-1}$&$\mathrm{km\,s}^{-1}$&$\mathrm{km\,s}^{-1}$&\\
\hline
thin disk &85$\pm$7&-0.17$\pm$0.03&0.26$\pm$0.03&-12$\pm$2&39$\pm$2&20$\pm$2&20$\pm$1\\
\hline 
thick disk&15$\pm$7&-0.48$\pm$0.05&0.32$\pm$0.03&-51$\pm$5&63$\pm$6&39$\pm$4&39$\pm$4\\
\hline 
\end{tabular}
\end{center}
\end{table*}

The proportion of thick disk stars is higher 
than expected in  the 2 bins of distance. If an
exponential law  with scale height of  800 pc is able  to modelise the
vertical density of  the thick disk, then a  local relative density of
15\% has to be adopted to  reproduce our results. On another hand, due
to the  small size  of the sample,  the proportions of  each component
cannot be recovered  with a high accuracy, as was also seen with the
simulated data. According  to the dispersion
of the SEM solutions on real and simulated data, we estimate our error
bar on local density to be 7\%.\par

Chiba \& Beers (2000) found
in their metal-poor star sample  a vertical gradient in the rotational
velocity  of the  thick disk  of  $-30\, \mathrm{km\,s^{-1}\,kpc^{-1}}$
which  is not  observed  in  the NGP  sample. We note however a higher 
$\sigma_W$ and lower [Fe/H] at 400 --  800\,pc possibly indicating a 
vertical gradient. As the metallicity found for the thick disk (-0.27 and -0.42)
is more metal-rich than expected, we have examined the eventuality of a bias  
in metallicity in our data. This high metallicity is mainly due to the lack
of stars with [Fe/H]\,$<$\,--0.65. As discussed in Sect. 3.2.4,
our method to estimate metallicities cannot explain by its own the complete 
absence of metal-poor stars.
The selection  of the target stars in the
colour range  $0.90 <  B-V < 1.10$  may have also led to a lack  of metal-poor
stars. We have investigated the effect of this colour cut on metallicity on data 
simulated with the Besan\c con model and on giants of the TGMET library. If
it is clear that deficient giants ([Fe/H]\,$<$\,--1.0) are bluer, the effect is not
so dramatic in the range --0.65\,$<$\,[Fe/H]\,$<$\,--1.0. Selecting
stars with $0.90 <  B-V$ seems to eliminate about  30\% of giants in this metallicity interval. 
Again, the lack 
of metal-poor stars cannot be completely explained by our colour cut. Besides, when
testing the SEM algorithm on simulated data (Sect. 4.3.2),  correct
mean metallicities  were recovered, despite a similar colour cut.
{\it If this lack of low metallicities
is real,  an explanation  would be that  the missing  metal-poor stars
belong  to another  population with  a higher  scale height  and lower
local density}. 
However this hypothesis does not explain the high metallicity, [Fe/H]=-0.27, found for the 
thick disk in  the nearest bin of distance. We interpret this value by 
the existence of stars
with slow  rotation but solar  metallicity  (Fig.~\ref{f:fehUVW}).  A  population of
metal-rich stars with  thick disk kinematics is known  to exist in the
solar  neighbourhood \citep{Gre99,So99}. This  population is  found to
have a  flat distribution in  the local disk  and its origin  near the
bulge, with   an   orbital  diffusion   by   a   bar,   is   proposed
\citep{MR99}.    These stars, which do not share the thick disk history,
can also    explain    the   lower
$\sigma_W$\,=\,28$\pm$3  \,$\mathrm{km\,s}^{-1}$ found  for  the thick
disk in the nearest  bin of distance, instead of $\sigma_W$=\,39$\pm$4
$\mathrm{km\,s}^{-1}$ farther from the plane. These considerations led
us to adopt  as the best parameters for the thick  disk the average of
the 4D and 3D SEM  solutions obtained  in  the  bin 400  --  800  pc. The  mean
metallicity of --0.42 is  probably still overestimated but yet favours
the  idea  of  a metal-rich  thick  disk.  Our  estimate of  the  mean
metallicity  of this  population is  [Fe/H] =  $-0.48 \pm  0.05$. The
adopted parameters of the thick disk are given in Table~\ref{t:defparam}. As
mentionned  previously,  numerous  determinations of  the  kinematical
parameters of the thick disk exist  in the literature and span a wide
range of values.  Our determination is also in this  range, but on the
cooler  side  with moderate  dispersions  in  V  and W,  and  moderate
asymmetric    drift.   On    the   contrary,    $\sigma_U    =   63\pm6
\mathrm{km\,s}^{-1}$  is significantly  higher than  most  of previous
determinations  which  are   often  below  $50\,  \mathrm{km\,s}^{-1}$
 \citep[e.g.][]{MFF90,So93,Ch97,CB00}.\par

We conclude  that our results  favour a dense, flat,  metal-rich thick
disk with moderate rotational and vertical kinematics, but significant
dispersion in  radial motions. Our  thick disk parameters  fit nicely
those found by \cite{Bell96} which also rely on a spectroscopic survey
of   570   K   giants,   but   in  the   rotation   and   antirotation
directions.  Similarly  to   us,  he  finds  a  lack   of  stars  with
[Fe/H]\,$<$\,-0.60. His sample is  better described by a two-component
disk model than by a continuous disk. He determines for the thick disk a
mean abundance of $-0.35 \pm 0.06$, an asymmetric drift of $-43\pm 9 \,
\mathrm{km\,s}^{-1}$, and a local normalization of $13 \pm 3 \%$. In terms
of formation of the thick disk, the discreteness and absence of a vertical
gradient suggest a quick heating of a precursor thin disk. This has to be
replaced in a context where a merger is at present the prefered model for the origin
of a thick disk in our galaxy and in other galaxies. The moderate parameters
that we have determined favour the interpretation that we have observed in
the NGP sample the puffed up early thin disk, whereas \cite{GWN02} and
\cite{CB00} have observed the relics of the disrupted satellite in the
form of slower rotating metal-poor population. The existence of two components
for the thick disk may reconcile the current inconsistent determinations of
its scale height and normalisation.

\section{Vertex deviation}
\label{s:ellipsoide}
\begin{figure}[hbtp]
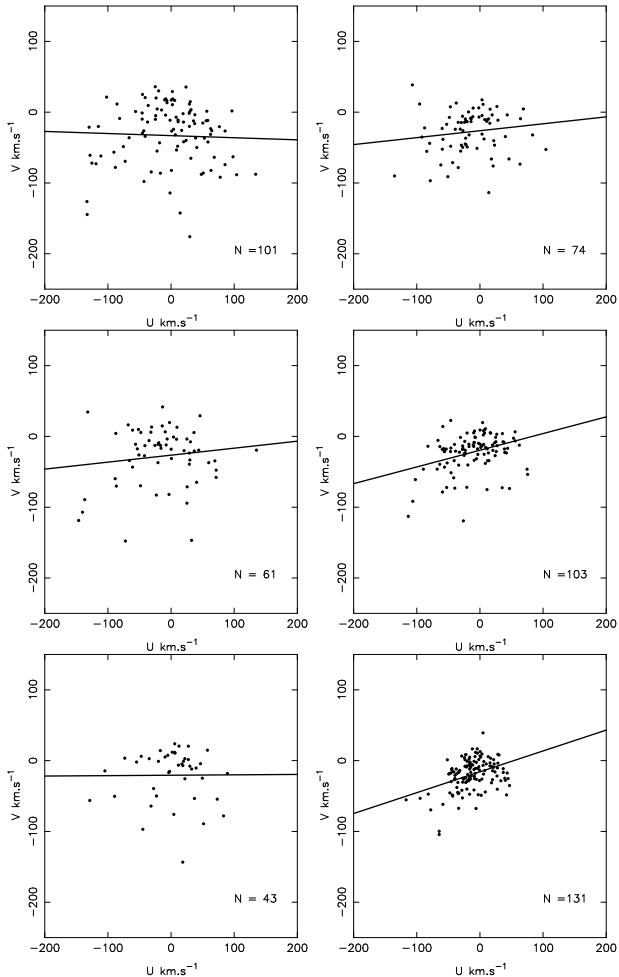
  
    \centering  
      \includegraphics[width=4.cm]{ps/UVpgn_mp_400+.ps}
      \includegraphics[width=4.cm]{ps/UVpgn_mr_400+.ps}
      \includegraphics[width=4.cm]{ps/UVpgn_mp_200-400.ps}
      \includegraphics[width=4.cm]{ps/UVpgn_mr_200-400.ps}
      \includegraphics[width=4.cm]{ps/UVhip_mp.ps}
      \includegraphics[width=4.cm]{ps/UVhip_mr.ps}  
      \caption{$(U,V)$  velocity distributions and vertex deviation. {\it  Left:} 
      selection  with [Fe/H]$\,
<\, -0.25$.  {\it Right:} selection  with [Fe/H]$\, >\,  -0.25$.  {\it
From       bottom       to       top:}       $z\,<\,125\,\mathrm{pc}$,
$200\,\mathrm{pc}\,<\,z\,<\,400\,\mathrm{pc}$,$z\,>\,400\,\mathrm{pc}$ } 
\label{f:ellipsoideUV} 
\end{figure}

The orientation of the velocity ellipsoid is of importance to test the
axisymmetry and  stationarity of the Galaxy.  The vertex deviation,
i.e.   the inclination  of  the  principal axis  with  respect to  the
direction of the Galactic Center is able to reveal distortions in the
disk as described in \cite{KT91}.  In order to follow the shape of the
velocity  ellipsoid with  distance to  the plane,  the NGP  sample was
combined  to   a  local  sample  constructed  out   of  the  Hipparcos
catalogue. Its  construction is described  in Paper II.  Briefly, this
sample consists  of mainly red  clump giants, complete up  to 125\,pc,
and   corrected   from  the Lutz-Kelker   bias.  Radial   velocities   and
metallicities  were  retrieved from  the  literature:  526 stars  have
radial velocities,  192 stars  have metallicities from  the literature
and 111  have metallicities derived from  Str\"omgren photometry.  The
NGP and  local samples have been  used to study the  variations of the
velocity ellipsoid with  respect to [Fe/H] and distance  to the plane,
and were divided into 6 subsamples: [Fe/H]$<-$0.25, [Fe/H]$>-$0.25 and
z\,$\le$\,125\,$\mathrm{pc}$,       200\,$<$\,z\,$<$400\,$\mathrm{pc}$,
z\,$>$\,400\,$\mathrm{pc}$.  The vertex deviation  is determined  by a
linear  fit in  $(U,\,V)$ for  each  subsample. Results  are presented  in
Table~\ref{t:orientationbis} and Fig.~\ref{f:ellipsoideUV}.\par

%
\begin{table}[t]
\caption{
\label{t:orientationbis}
For each subsample, orientation in
degree  of   the   velocity   ellipsoid  from a linear regression on the ($U,\,V$) distribution.
Errors range from $3\degr$ to $5\degr$.}
\begin{center}
\begin{tabular}{c c c}
\hline
\hline
Selection & [Fe/H]$\,<\,-0.25$ & [Fe/H]$\,>\,-0.25$ \\
\hline
$z\,\geq\,400\,\mathrm{pc}$ & -1.7$\degr$ & 5.6$\degr$ \\
$200\,\mathrm{pc} \,<\,z\,\leq\,400\,\mathrm{pc}$ & 5.6$\degr$ & 13.3$\degr$ \\
$z\,\leq\,125\,\mathrm{pc}$ & 0.3$\degr$ & 16.4$\degr$ \\
\hline
\end{tabular}
\end{center}
\end{table} 
A positive  vertex deviation (Fig.~\ref{f:ellipsoideUV})  is seen in
the high metallicity sample in the  3 bins of distance, with however a
smaller deviation in the highest  bin. Kinematical groups are visible in
the nearest bin of metal rich stars, suggesting a younger, partially mixed  
population. If we consider the inclination
of the inner part of the  ellipsoids, the vertex deviation is found to
range between  $\sim13\degr$ and  $\sim25\degr$. On the  contrary, the
low metallicity  sample, dominated by  thick disk stars, has  a vertex
deviation  near zero.   We may  also remark  that metal rich  stars  
have velocity dispersions and vertex deviation intermediate between old  
disk  and and  thick  disk in the highest bin.\par

It  is well  established and  recently confirmed  from  Hipparcos data
\citep{DB98, B99}  that the  vertex deviation declines  from $25\degr$
for young stellar populations to near zero for older and kinematically
hotter disk populations.  This deviation of the vertex orientation has
been  frequently  discussed  \citep{KT94,DB98,BM98}  and  may  imply
either a non-axisymmetric Galaxy or local spiral perturbations or more
simply partially mixed populations.  Since the vertex deviation is not
constant, the decline suggests that the deviation is caused by a small
scale peturbation  like a nearby  spiral arm or  by non-stationarity,
the deviation  reflecting the recent  condition of formation  of young
stars. From Hipparcos data, \cite{B99} finds that the vertex deviation
is compatible with zero (5$\pm5\degr$) for the reddest Hipparcos stars
($B$-$V\sim$\,0.5--0.7),  while  \cite{DB98}  obtain a  10$\pm5\degr$
deviation for the same stars.  From their respective result, they draw
opposite  conclusions concerning the  axisymmetry or  non-axisymmetry of
the galactic  disk, but  we may remark  that their  measurements agree
within $1\sigma$ error.\par

Here we find that the  metallicity allows a more drastic separation of
populations, and  that the  oldest and well-mixed population  has a
null vertex  deviation. $B-V$  colours do not  allow to  separate very
efficiently Hipparcos  populations by age and the  red Hipparcos stars
are a mixture of old and a small proportion of young stars producing a
non strictly null vertex deviation. This is clearly established, here,
with the Hipparcos sample of giants with known metallicities.\par

In  conclusion  the vertex  orientation  of  the  oldest stars  points
towards  the   galactic  center,   favouring  the  hypothesis   of  an
axisymmetric galactic  disk (however if the galactic  disk is elliptic
and the  Sun is located  on one of  the symetry axis, the  vertex will
also point towards the Galactic Center).\par

%
\section{Conclusion}
We have presented a new sample of nearly 400 stars selected out of the
Tycho--2 catalog and observed at high spectral resolution. This sample
mainly  consists of  red clump  giants.  Their  atmospheric parameters
(T$_{\rm eff}$, $\log g$,  [Fe/H]) and absolute magnitudes $M_{\textrm
v}$  were  estimated  with  the  TGMET software  by  comparison  to  a
reference library of  spectra. The sample was used  to investigate the
kinematics and metallicity  of the stellar disk up to  800 pc from the
galactic   plane  with  a   typical  error   of  18\%   on  distances,
$3\,\mathrm{km\,s}^{-1}$  in  U  and  V  velocities,  and  lower  than
$1\,\mathrm{km\,s}^{-1}$  in W,  at the  mean distance  of  the sample
(400\,pc).\par

Our colour-magnitude cuts were efficient to select K giants since less
than  6\% of  the  sample consists  in  dwarfs or  subgiants. We  have
observed    in   the    sample    a   lack    of   metal-poor    stars
([Fe/H]\,$<$\,-0.65) which is  only partially explained by
the method used to estimate the metallicities and  by the colour cut.
Velocity  and
metallicity distributions were compared to simulations of  the Besan\c con
model using 2 models of thick disk: a metal-poor kinematically hot one
and  another one  with parameters closer to  the thin  disk.  The NGP 
sample is in a better agreement with the moderate thick disk model. Using  
a non-informative
method in the  case of a mixture of gaussian  populations, we find our
sample to be  consistent with a superposition of  two disks. The first
one,  associated  the   thin  disk,  has
$V_{\rm lag}=-12\pm2\, \mathrm{km\,s}^{-1}$ with respect to the Sun, ($\sigma_U,
\sigma_V,  \sigma_W)=(39\pm2,20\pm2,20\pm1)\,\mathrm{km\,s}^{-1}$,
[Fe/H]\,=\,--0.17\,$\pm$\,0.01,  in  agreement  with previous  studies
from  local  samples.   No   vertical  gradient  is  observed  in  the
parameters of the thin disk. \par

 The  parameters  of  the  thick  disk were
 estimated to be $V_{\rm lag}=-51\pm5\,\mathrm{km\,s}^{-1}$,
($\sigma_U,  \sigma_V,  \sigma_W)=(63\pm6,39\pm4,39\pm4)\,\mathrm{km\,s}^{-1}$,
 [Fe/H]\,=\,--0.48\,$\pm$\,0.05.  Compared  to previous studies, these
 values  are on the  hot side  for U,  but on  the cold,
 metal-rich side  for V,  W and [Fe/H].  The proportion of  thick disk
 stars is found to be  higher than expected with a local normalisation
 of 15$\pm$7\%. Inconsistent determinations of the thick disk parameters in
 the recent literature may be reconcilied if 2 components are involved. In the
 context of the formation of the thick disc by a significant merger, we
 may have observed in the NGP sample the early heated thin disk, whereas other 
 authors may have observed the relics of a shredded satellite.\par

The vertex deviation  of the velocity ellipsoid has  been studied with
respect to  distance above the  plane and metallicity. For  this task,
the NGP  sample was  completed by a  local Hipparcos sample  for which
metallicity   and   radial   velocities   were  retrieved   from   the
literature. A null  vertex deviation is found for  the low metallicity
subsample whereas the high  metallicity sample exhibit a positive one.
This  suggests no strong  deviation from  axisymmetry of  the galactic
disk at solar radius.\par

The next paper of
this series will  give a new derivation of the mass  density in the galactic
plane based on these observations  combined to the local  Hipparcos sample.
The sample presented in this paper, together with its Hipparcos counterpart,
provides  a  sample  with  unprecedented  accuracy  to  study  the  vertical
potential up to 800~pc.
%
\begin{acknowledgements}
This  research  has made  use  of  the  SIMBAD and  VIZIER  databases,
operated at CDS, Strasbourg,  France. It is
based  on data from  the ESA {\it  Hipparcos} satellite
(Hipparcos and Tycho--2 catalogues). We thank 
A. Robin who provided simulated samples from of new version of the Besan\c con model
before publication.

\end{acknowledgements}
%
%
%

\end{document}